\documentstyle[preprint,aps]{revtex}
\begin{document}
\draft
\title{New 
Evidence for ``Confined Coherence'' in Weakly Coupled Luttinger Liquids}
\author{David G. Clarke$^1$\cite{byline_me} and S. P. Strong$^2$
\cite{byline_steve}}
\address{$^1$IRC in Superconductivity and Cavendish Laboratory, 
University of Cambridge,\\
Cambridge, CB3 0HE, United Kingdom \\
$^2$NEC Research Institute, 4 Independence Way,
Princeton, NJ, 08540, U.S.A.\\}
\date{May 2, 1996; revised January 8, 1997}
\maketitle
\begin{abstract}
Based upon the calculation of
the exact interliquid hopping rate and an approximate single particle
Green's function, we present new evidence for the existence of a
phase of {\em relevant but incoherent\/} inter-Luttinger liquid transport.
This phase of ``confined coherence'' occurs
when the Luttinger liquid exponent $\alpha$ satisfies $\alpha_c<\alpha<1/2$.
We argue that $\alpha_c$ is strictly bounded above by $1/4$, 
and is probably substantially
smaller, especially in spin-charge separated Luttinger liquids.
We also discuss connections with the work of others. ~\\

\end{abstract}




\section{Introduction}
The physics of a strongly correlated, highly anisotropic electron
system represents a subtle problem in many-body physics.
In previous work \cite{prl,lees2} we have considered the
problem of one-dimensional (1D) electron liquids coupled by 
weak interliquid hopping. Implicit in our approach is the
recognition that if one begins
with a collection of truly 1D metals and then turns on weak interliquid hopping,
it is not {\em a priori\/} appropriate to consider the electron-electron
interaction as a perturbation on an anisotropic (2D) free Fermi gas.
Rather, one should consider the {\em interliquid hopping\/} as a perturbation
on the (otherwise decoupled) 1D liquids.
The problem is nontrivial
but tractable to some degree because
the low energy physics of a 1D metal
is described by Luttinger liquid
theory.
Unlike in a Fermi liquid, where the
electron spectral function, $\rho(k,\omega)$
is dominated by a quasiparticle part, which sharpens up to a 
$\delta$-function as $k\rightarrow k_F$, in a Luttinger liquid
there are no Landau quasiparticles, rather, $\rho(k,\omega)$ exhibits
only power law singularities. For this reason,
and others we have previously discussed \cite{prl,lees2},
the problem of weakly coupled Luttinger liquids is closely analogous 
to that of weak tunneling in a two level system (TLS) coupled to
an ohmic dissipative bath \cite{TLS}. Exploiting this analogy
led us to propose \cite{prl,lees2} that interliquid hopping
between  non-Fermi liquids
may have three qualitatively distinct
regimes: it may be irrelevant, 
relevant and coherent, or relevant but entirely incoherent.
The incoherent interliquid hopping phase 
would represent a new state of 
matter 
with intrinsically incoherent transport in at least one direction.
There is substantial
experimental support
for this proposal based on its
ability to explain certain anomalous properties
of the low dimensional organic conductor (TMTSF)$_2$PF$_6$,
as has been discussed elsewhere
\cite{magic,guynfl,biggie}.
In this 
paper we briefly report
new results which address this question
based upon the use of exact Luttinger liquid spectral functions,
careful consideration of the analytic properties
of Luttinger liquid Green's functions, and
a reinterpretation of a calculation made by others \cite{castellani}.

\section{Interliquid Hopping Rate for Weakly Coupled Luttinger Liquids}
We are interested in the problem of $N$ coupled Luttinger liquids, 
$N\rightarrow\infty$. At $O(t_{\perp}^2)$, however, our results are
equivalent to those for $N=2$, and we therefore consider
the problem of two Luttinger liquids coupled
by a spatially uniform, single particle hopping (as in \cite{prl}).
Our 
calculation is  dynamical and
involves taking a $t=0$ state with $\Delta N$ more
(right moving) particles in liquid 2 
than liquid 1 and no Tomonaga bosons excited in
either, then turning on $t_{\perp}$
and examining
the time dependence of $\Delta N$
(for motivation see \cite{prl} and \cite{TLS}). The particle number
difference $\Delta N$ entails a Fermi momentum difference $\Delta k$
and a chemical potential difference $\Delta\mu\equiv v\Delta k$.
Unlike our earlier work \cite{prl} based upon space-time Green's functions,
we use spectral function methods here,
which is both physically more illuminating and permits the 
calculation of key correlation functions {\em exactly\/}.

At $O(t_{\perp}^2)$ the {\em interliquid hopping rate\/}
$\Gamma(t)$
can be written in a spectral function
form as
\begin{equation}
\Gamma(t)=2t_{\perp}^2L\int\frac{d\omega}{2\pi}
\frac{\sin\omega t}{\omega}\{A_{12}(\omega)+A_{21}(\omega)\}
\label{eq: gamma(t)}
\end{equation}
where
\begin{equation}
A_{ij}(\omega)=\int\frac{d\omega'}{2\pi}
\int\frac{dk}{2\pi}
{\cal J}_1^{(i)}(k,\omega'){\cal J}_2^{(j)}(k,\omega'-\omega)
\label{eq: A's}
\end{equation}
and the spectral functions ${\cal J}_{1,2}(k,\omega)$ are the Fourier transforms of
${\cal J}_1(k,t)\equiv\langle c_1(k,t)c_1^{\dag}(k,0)\rangle$ and
${\cal J}_2(k,t)\equiv\langle c_2^{\dag}(k,0)c_2(k,t)\rangle$.
In this paper we consider only the zero temperature limit, in which case
${\cal J}_{1,2}(k,\omega')=
\theta_{\pm}(\omega'-\mu)\rho_{1,2}(k,\omega'-\mu)
$
where $\rho(k,\omega)$ is the electron spectral function as conventionally
defined.
We remark that Eqs. (\ref{eq: gamma(t)}) and (\ref{eq: A's}) are not
specific to coupled 1D liquids: they may be extended to the case of coupled
2D liquids 
by replacing $k$ by ${\bf k}$ in the $k$-integrals and in the definitions of
${\cal J}_1$ and ${\cal J}_2$.

Physically, $A_{12}(\omega)$ is the effective interliquid hopping
spectral function
for an electron hopping {\em to\/} liquid 1, {\em from\/} liquid 2,
and $A_{21}(\omega)$ the opposite. 
As $A_{21}(\omega)$ never has a
coherent component, it suffices,
for the purposes of studying the question
of coherence, 
to consider only $A_{12}(\omega)$.

Before presenting the calculation of $\Gamma(t)$ for coupled Luttinger
liquids, we first show how the coherence of interliquid hopping
manifests itself in the case of coupled (Landau) Fermi liquids.

{\em Free Fermi Gasses, and Fermi Liquids:}
For free Fermi gasses, $A_{12}(\omega)\propto\Delta\mu\delta(\omega)$ and
$A_{21}(\omega)=0$. Thus $\Gamma(t)\propto\Delta\mu~t$, a clear signal of
coherent hopping and hence of a fundamental rearrangement of the ground
state.

In a Fermi liquid 
the (retarded) Green's function is
$G_R^{-1}(k,\omega)=Z^{-1}(\omega-E_k)+i\gamma\omega^2$
where $Z$ is the quasiparticle 
renormalization factor, and $\gamma$ is a (positive) parameter
characterizing the strength of the electron-electron interactions.
The spectral function is then given by 
$\rho(k,\omega)=-2\;{\rm Im}G_R(k,\omega)$
from which we obtain
\begin{eqnarray}
A_{12}(\omega)&\sim&v_F^{-1}
\{Z^2\Delta\mu\delta(\omega)+(3\pi)^{-1} Z^3 \gamma \omega\}
\,\theta_+(\omega+\Delta\mu) 
\end{eqnarray}
We find that
$\Gamma_{12}(t)$ is a sum of a term $\propto Z^2\Delta\mu t$
representing fundamentally coherent processes, and a term 
$\propto\gamma Z^3 t^{-1}$ which is marginal.
By choosing a sufficiently small $t_{\perp}$ one can
find a time $t$ such that, 
while remaining in the perturbative regime, $N^{-1}\int_0^t\Gamma(t')dt'\ll 1$,
the ratio of the coherent contribution to the marginal contribution
is arbitrarily large. This is true {\it regardless of how small $Z$ is\/}.
Thus, a perturbative calculation in $t_{\perp}$ does not reveal any
likelihood of a loss of coherence of interliquid tunneling, and there is
no impediment to the formation of an interliquid band of
width $\sim Zt_{\perp}$. 
This is consistent with
what we would expect from a calculation based upon (Landau) quasiparticles.
Formally, the coherence is reflected in the fact that the spectral
function $A_{12}(\omega)$ is dominated by the $\delta$-function at
$\omega=0$, indicating that hopping is almost entirely energy degenerate.

{\em Luttinger Liquids:}
We now turn to the problem of coupled Luttinger liquids, considering
the case of spin-independent
electronic interactions, characterized by the 
anomalous exponent, $2 \alpha$, of the single particle 
Green's function,
and charge- and spin-velocities $v_c$ and $v_s$.
The calculation of $A_{12}(\omega)$
and $A_{21}(\omega)$ is lengthy, and we present only the final results here.
Complete details are given in \cite{biggie}.
The {\em exact\/} result is 
\begin{eqnarray}
A_{12}(\omega)&=&A_{12}^{\rm low}(\omega)\,
\theta_+[\omega-(v_s-v)\Delta k]\,\theta_+[(v_c-v)\Delta k-\omega]
+
A_{12}^{\rm high}(\omega)\,
\theta_+[\omega-(v_c-v)\Delta k] \nonumber \\
A_{12}^{\rm low}(\omega)&=&\frac{1}{\Gamma(1+4\alpha)}
\frac{1}{\Delta v}
\left(\frac{a^2}{\bar{v}\Delta v}\right)^{2\alpha}
(\omega+(v-v_s)\Delta k)^{4\alpha} \label{eq: LL eff spectral} \\
A_{12}^{\rm high}(\omega)
&=&\frac{1}{(1+2\alpha)}\frac{1}{\Gamma(2\alpha)\Gamma(1+2\alpha)}
\frac{1}{\bar{v}} 
\left(\frac{a}{2v_c}\right)^{4\alpha}
(\omega+(v_c+v)\Delta k)^{2\alpha+1}(\omega-(v_c-v)\Delta k)^{2\alpha-1}
\nonumber \\
&&_2F_1\left(1, 1-2\alpha;2+2\alpha;-\left(\frac{\Delta v}{\bar{v}}\right)
\left[\frac{\omega+(v_c+v)\Delta k}{\omega-(v_c-v)\Delta k}\right]\right)
\nonumber
\end{eqnarray}
where $_2F_1$ is the hypergeometric function, $a$ a short distance cutoff,
and $\bar{v}\equiv v_c+v_s$, $\Delta v\equiv v_c-v_s$ \cite{footnote2}.
The typical morphology of
$A_{12}(\omega)$ is shown in Fig. 1.
We observe that $A_{12}(\omega)$ is both nonsingular and of wide support,
having non-zero weight from just below $\omega=0$ all
the way up to the ultraviolet cutoff. 
As $\alpha\rightarrow 0$, we have $A_{12}(\omega)\rightarrow\delta(\omega)$,
and one needs to use degenerate perturbation theory to treat
the interliquid hopping. For $\alpha > 1/2$, $t_{\perp}$ is a 
formally irrelevant operator, which is reflected in the fact that 
the spectral weight in $A_{12}(\omega)$ is pushed to high energies.
For $\alpha < 1/2$, but not too small, $A_{12}(\omega)$ is generically
``flat'' suggesting that much, if not most,
 of the hopping occurs via non-degenerate
({\em i.e.\/} ``inelastic'') processes. This is reminiscent of 
situations in more elementary quantum mechanical problems where 
Fermi's ``Golden rule'' is applied, and clearly raises doubts over any claim
that the action of $t_{\perp}$ is to drive the system to a fixed point
in which interliquid hopping is coherent.

For simplicity, we shall restrict our discussion from here on to
the spinless case, which can be obtained by formally taking $\Delta v\rightarrow 0$.
The general case shall be discussed elsewhere \cite{biggie}.

In calculating $\Gamma_{12}(t)$ it is simplest to consider
its time derivative.
We find
\begin{eqnarray}
\frac{d\Gamma_{12}(t)}{dt}
&=&\frac{t_{\perp}^2 L}{\pi}
\frac{1}{\Gamma(2\alpha)\Gamma(2+2\alpha)}
\left(\frac{a}{2v_c}\right)^{4\alpha} 
\frac{1}{2v_c}\frac{1}{\Gamma(1-2\alpha)}t^{-(1+4\alpha)} \nonumber\\
&&{\rm Re}\{
e^{i(v_c-v)\Delta kt}
[ie^{i2\pi\alpha}\Gamma(1-2\alpha)\Gamma(1+4\alpha)~_1F_1(-1-2\alpha,-4\alpha;-ix)
\nonumber \\
&&
+\frac{1}{2}\frac{(1+2\alpha)}{(1+4\alpha)}
\Gamma(2\alpha)\Gamma(1-4\alpha)\;
x^{1+4\alpha}~_1F_1(2\alpha,2+4\alpha;-ix)
]\}
\label{eq: spinless exact}
\end{eqnarray}
where $~_1F_1$ is the confluent hypergeometric function and, 
for convenience, we have introduced the variable $x=2v_c\Delta k t$.

Equation (\ref{eq: spinless exact}) is an {\em exact\/} result for
the (time derivative of the) interliquid hopping rate, to lowest
order in $t_{\perp}$.
We use the expansion 
$~_1F_1(a,b;z)=1+ab^{-1}z+ab^{-1}(a+1)(b+1)^{-1}z^2/2
+\ldots$
and, noting that
it makes little physical sense to suppose that terms of 
$O(x^2)$ or higher ({\em i.e.\/} terms of $O(\Delta k^2)$ or higher)
are important
in determining the coherence or incoherence of {\em single particle\/} hopping,
we retain only the $O(x^0)$, $O(x)$ and $O(x^{1+4\alpha})$ terms.
This gives
\begin{eqnarray}
\frac{d\Gamma_{12}(t)}{dt}
&=&\frac{t_{\perp}^2 L}{\pi}
\frac{1}{\Gamma(2\alpha)\Gamma(2+2\alpha)}
\left(\frac{a}{2v_c}\right)^{4\alpha} 
\frac{1}{2v_c}\,\cos[(v_c-v)\Delta k t]\,t^{-(1+4\alpha)} \nonumber\\
&&\{(1+2\alpha)
\left\{-\sin(2\pi\alpha)\frac{\Gamma(1+4\alpha)}{(1+2\alpha)}
+\cos(2\pi\alpha)\Gamma(4\alpha)\,x
+\frac{\Gamma(2\alpha)\Gamma(1-4\alpha)}{2(1+4\alpha)\Gamma(1-2\alpha)}\,
x^{1+4\alpha}\right\} \nonumber \\
&&-\tan[(v_c-v)\Delta k t]\,\cos(2\pi\alpha)\Gamma(1+4\alpha)\}
\label{eqn: spinless gamma dot}
\end{eqnarray}
The latter two terms continuously develop into the 
correct, coherent result for free fermions
as $\alpha\rightarrow 0$ and
their modification from the Fermi liquid
result
is closely analogous to the behavior of the appropriate 
terms in the derivative of the
TLS transition rate upon turning on
coupling to an ohmic bath.
In the other 
well understood limit, $\alpha > 1/2$,
the entire expression
leads to a finite integrated 
transition probability, $P(t) = \int_0^{t} dt' \Gamma(t')$,
in agreement with the known 
irrelevance of $t_{\perp}$ in that limit.
In between, the
$\Delta k$-independent term gives 
$P(t) \propto t^{1-4\alpha}$ which is
long-time convergent and therefore 
irrelevant if $\alpha > 1/4$, but
represents fundamentally incoherent
interliquid hops if $0~<~\alpha~<~1/4$. 
The $O(\Delta k^{1+4\alpha})$ term requires care to interpret
when $\alpha\neq 0$, but we note that the oscillatory prefactor
$\cos[(v_c-v)\Delta k t]$ will force $\Gamma_{12}(t)$ to be essentially
time-independent for times $t~^>_{\sim}[(v_c-v)\Delta k]^{-1}$.
This effect
is analogous to that of non-degeneracy in the TLS
\cite{TLS} where it has been argued to dramatically
decrease coherence. 
In order for 
lowest order hopping
to be coherent, one must remain at times short enough to avoid the
cutoff effect of 
this prefactor and the  maximum
possible $\Delta k$ for a given time $t$ is 
$\Delta k^{\rm max}\sim [(v_c-v) t]^{-1}$. The $O(\Delta k^{1+4\alpha})$ term
in $d\Gamma/dt$ is therefore bounded by 
$\sim \Delta k\,t^{-4\alpha}/(v_c-v)^{4\alpha}$ which has the same form
as the term linear in $\Delta k$ and we therefore consider only
the latter term.

If the term linear in $\Delta k$ decays slower than $t^{-1}$
it should be interpreted
as a potentially coherent term.
For $\alpha > 1/4$ 
it falls off faster than $t^{-1}$
and at
$O(t_{\perp}^2)$
the interliquid single particle hopping
is completely incoherent, signalled by the
convergence of $\Gamma(t\rightarrow\infty)$.
{\em This is despite the relevance of 
$t_{\perp}$ in the RG sense for $\alpha < 1/2$\/}. 
We therefore expect an 
incoherent interliquid hopping phase for $1/4 < \alpha < 1/2$.
There are, however, additional
factors enhancing incoherence over and above the time exponent of
the $O(\Delta k)$ term.

First, 
there is the ``dephasing'' prefactor
$\cos[(v_c-v)\Delta k t]$, analogous to a bias term in a TLS.
Based on the results
from that problem \cite{TLS}, this should enhance incoherence.
Further,
there are
the incoherent processes contributing to the $\Delta k$-independent term.
For $0 < \alpha <1/4$ the interliquid hopping rate 
and
the integrated
transition probability, $P$,
are essentially
sums of incoherent and coherent parts, 
defined by their respective time behaviors.
Due to the presence of the dephasing term, the coherent term remains so
only
for times $t~^<_{\sim}[(v_c-v)\Delta k]^{-1}$. As such, $P_{12}^{\rm coh}(t)$
is bounded above in magnitude by 
$\sim t_{\perp}^2 v_c \Lambda ^{4\alpha} t^{1-4\alpha}/(v_c-v)$ so that,
\[
\frac{P_{12}^{\rm incoh}(t)}{P_{12}^{\rm coh}(t)}~
^{>}_{\sim}\,\alpha\frac{(v_c-v)}{v_c}
\]
This is {\em independent\/} of $t_{\perp}$
and the purely incoherent channel cannot be eliminated in the 
$t_{\perp}\rightarrow 0$ limit, as it can in a Fermi liquid.
As a result, we are forced to consider
the influence of interliquid hops upon one another via
correlations
not automatically included in our
$O(t_{\perp}^2)$ calculation. 
To begin with, interliquid hops through the coherent channel will be
interrupted by the finite probability of a hop through the incoherent
channel. Secondly, intraliquid interactions will lead to scattering of 
coherent hops by incoherent hops.
In the limit $t_{\perp}\rightarrow 0$,
the incoherent hops have an arbitrarily long time
to scatter the coherent hops
(although their density also vanishes in this limit)
and we find that  
the effect of a given incoherent hop on the coherent hops 
grows at least linearly in time.
If it grows faster than linearly, 
the scattering will be divergent and
hopping should be incoherent
as $t_{\perp}\rightarrow 0$ {\em for any\/} $\alpha$. 

Combining all of these effects,  we expect that
as we decrease $\alpha$ from $1/4$,
incoherence will be stabilized down to some critical value $\alpha_c<1/4$
by a combination of the purely incoherent term,
the  dephasing prefactor 
which kills coherence if $\Delta k t$ is too large,
and, in the case of fermions with spin, 
spin-charge separation,
which further suppresses coherence for finite $\Delta k t$
\cite{biggie}.
We again emphasize the utility of the spectral function
$A_{12}(\omega)$ in indicating the coherent or
incoherent nature of the interliquid hopping.

\section{Approximate Single Particle Green's Function: 
Calculation and Interpretation}
We now consider 
how these same effects might appear
in the more conventional calculation
of the Green's function for $N\rightarrow\infty$ coupled Luttinger liquids of
spinless fermions.
We will neglect
vertex corrections
associated with $t_{\perp}$ and
incorporate
$t_{\perp}(k_{\perp})$ as an energy
independent self-energy.
We are motivated by similar calculations by 
others \cite{wen}, however
we focus on analytic properties
of the Green's function not previously treated.
Using $G^{-1}=G_0^{-1}-\Sigma$, 
$G_0(k,\omega)=(v^2 k^2-\omega^2)^{\alpha}(\omega-vk)^{-1}$,
 gives
\begin{equation}
\label{eq:G}
G(k,k_{\perp},\omega) =
\frac{(v^2 k^2 - \omega^2)^{\alpha}}
{(\omega-vk)-t_{\perp}(k_{\perp})(v^2 k^2 -\omega^2)^{\alpha}}
\end{equation}
where we have set the dimensionful
high energy cut-off to $1$ and $k$ is 
momentum along the chains measured
from the Fermi surface.
Eq. \ref{eq:G}  must be supplemented by a 
discussion of the analytic properties of
$G$ and $G_0$, for whose discussion
we consider 
only positive $k$ since
negative $k$ is essentially identical
for $t_{\perp}(k_{\perp}) \rightarrow -t_{\perp}(k_{\perp})$. 

First, recall that the singularities of the Green's function,
particularly poles, 
only have sensible physical interpretations
in the second and fourth quadrants of the complex $\omega$
plane.  For $k \neq 0$, 
$G_0$ has
two branch cut singularities, one for each sign of $\omega$,
which must originate in the second and fourth quadrants.  Also,
$G_0$ must be real
for $-vk < \omega < vk$ since in that region no on-shell decay
of an injected fermion is possible.
This implies that 
the phase of $G_0$ for $\omega > v k$
should be given by $-\alpha \pi$, and
by $-\pi - \alpha \pi$ for $\omega < -vk$.
Now consider 
the pole equation,
$G_0^{-1}(k,\omega)=t_{\perp}(k_{\perp})$,
for $k=0$ and $t_{\perp}(k_{\perp})>0$.
For $\alpha = 0$,
the pole in the
complex $\omega$ plane
is at $t_{\perp}(k_{\perp})$
and, as we turn on $\alpha$, it shifts into the
fourth quadrant.
Moving off the axis into the
fourth quadrant an angle $\Theta$ changes the phase
of $G^{-1}_0(0,\omega)$
to  $\alpha \pi - (1-2\alpha) \Theta$ and it is again possible
to have a pole if
\begin{equation}
\Theta = \alpha\pi/(1-2\alpha)
\end{equation}
For small $\alpha$, this pole could be sensibly interpreted
as a weakly damped quasiparticle pole, as in a usual
Fermi liquid.
However, for $\alpha > 1/4$, $\Theta > \pi/2$, the pole
enters the fourth quadrant
and the solution has no
sensible interpretation as a quasiparticle
pole (it would imply an unoccupied, negative energy
quasiparticle state).  The last physical solution,
which occurred for $\alpha = 1/4$, corresponds to a
purely imaginary frequency, entirely in keeping with the idea
that $t_{\perp}$ is  acting incoherently at this value
of $\alpha$.
For a negative $t_{\perp}$, an exactly parallel scenario
involving the second, instead of the fourth quadrant,
results. In both cases, for $\alpha > 1/4$,
there is no physically sensible pole resulting from
incorporation of
$t_{\perp}$ as a self-energy, and the results are extremely
suggestive of incoherence.

The effect is very closely analogous to the behavior of the
Laplace transform of $P(t)$ found in \cite{TLS}
at the onset of incoherence.   A similar
analogy between the locations of the poles of
the single particle Green's function approximated in
this way and the Laplace transform
of $P(t)$ in the TLS problem was noted in
\cite{sudip}.


We now consider $k \neq 0$.
Consider first the case $t_{\perp}>0$.
As we move some distance away from the Fermi surface,
the singularity at $-v k$ becomes more distant and
its effect on the phase less important.
For $k^{1-2\alpha} \gg t_{\perp}$,
it becomes possible to circle the singularity at
$vk$ without moving appreciably with respect to
the singularity at $-vk$,
and the phase of $G_0^{-1}(k,\omega)$
close to $\omega = vk$ varies as
$\alpha \pi - (1-\alpha) \Theta$
where $\Theta$ is
measured downward
from the real $\omega > vk $ half-line.
As before, at small $\alpha$ the pole has a small imaginary
part to its frequency and it can be sensibly interpreted
as a weakly dampled quasiparticle pole.
However, now $\alpha$ can be as large as $1/2$ before the pole
is forced into an unphysical region.
Note, however, that for $\alpha > 1/3$, $\Theta > \pi/2$, so that
the addition of a positive, real self energy ($t_{\perp}$) shifts
the singularity at $vk$ to a complex energy with a real part 
{\em less than\/} $vk$.
Including spin-charge separation
is more complicated and we state only two 
of our results for $v_c>v_s$:
(1) for $\alpha > 1/6$, the pole lies
at an energy whose real part is shifted  lower
than $v_{c} k$, and (2) for $\alpha > 1/4$,
and large $k$ the pole equation is again not satisfiable
for a physically sensible $\omega$.

Returning to the spinless case, let us
follow the pole for $t_{\perp}<0$
as we increase $k$.  For $\alpha< 1/4$ this pole lies
in the second quadrant
and
increasing $k$ eventually 
pushes it to the imaginary axis.
When $k=k_{c} = v^{-1} t_{\perp}^{1/(1-2\alpha)} \cos(2\pi\alpha)$,
the pole reaches
$\omega = \omega_c = i\,v k_{c}  \tan(2\pi\alpha)$.
Again, the last frequency with a possible
physical interpretation is purely imaginary,
paralleling what occurred for $k=0$ and $\alpha = 1/4$.

In addition to this pole,
a new pole appears, for negative $t_{\perp}(k_{\perp})$
and $k>0$, at $\omega \in (-vk, vk)$ given by the real solution of
$(vk-\omega)^{1-\alpha}(vk+\omega)^{-\alpha}=|t_{\perp}(k_{\perp})|$.
This undamped pole is unphysical, however, in a number of ways.
Firstly, $G$ is purely real at the position of the pole
only because there
is nothing for a fermion at this momentum and
energy to decay into in the unperturbed model.
It is easy to see, however, that if the pole existed for
$k$ close to $0$, then there would be accessible decay
channels.  These are neglected by the omission of
vertex corrections. Secondly, for $|t_{\perp}(k_{\perp})|>2vk$, this pole
approaches not $vk-t_{\perp}$
but $-vk$ (with rapidly vanishing weight)
as $\alpha\rightarrow 0$. Finally, in a spin-charge separated Luttinger
liquid, and at sufficiently large $k$, 
this pole ceases to exist if $\alpha >1/4$,
while for $\alpha< 1/4$ the pole lies just below $v_{s}k$.
In a model with $\alpha< 1/4$ and vanishing spin velocity,
e.g. the large $U$ Hubbard model,
the pole is completely dispersionless
along the chains. The ``quasiparticles'' defined
by it have the strange property of a vanishing bandwidth
in the direction of large hopping, but
a finite bandwidth
in the direction of small hopping!

We therefore see that, for $\alpha \geq 1/4$
and when analytic properties are treated
carefully, 
this approximate calculation of $G$
gives no indication of the existence of
a transversely dispersing quasiparticle.
This is in contrast to
what has been suggested elsewhere
\cite{wen}.  In fact, if the poles found off the real 
axis are interpreted
as quasiparticle poles and the Fermi surface is identified
with the momenta at which the real part of their frequencies
cross zero energy, then the conclusion within this approximation
is that the Fermi surface warping vanishes completely for
$\alpha = 1/4$.

\section{Another Diagrammatic Calculation: Fermi Surface 
Warping}
Finally, we briefly discuss another diagrammatic
calculation addressing coupled Luttinger liquids.
For the case of infinitely many coupled chains
the behavior of $n(k_x,k_y)$
has been
studied in lowest order perturbation theory
in $t_{\perp}$
by Castellani {\it et al.} \cite{castellani}.
They find a shift of $n(k_x,k_y)$ 
proportional to
$\cos(k_y) |k^F_x - k_x|^{-1+4\alpha}$
and interpret this as signaling the instability
of the Luttinger liquid.  The $k$ behavior
arises from an integral given in 
our language by:
\begin{equation}
\langle \delta n(k_x,k_y) \rangle \propto
\cos(k_y)
\int \frac{d\omega}{2\pi} \frac{A^{\Delta N = 0}(k_x,\omega)}{\omega}
\end{equation}
which is {\it infrared convergent everywhere}  
for $\alpha > 1/4$.

We have previously
argued \cite{prl} that the magnitude of the warping of
the Fermi surface
should be identified with the
oscillation frequency of 
our dynamical calculation and
provides the order parameter
for the transition between the phase with
``confined  coherence'' and
the usual phase with coherent transport in
all directions. 
When interpreted in this context,
the finding of
Castellani {\it et al.} \cite{castellani},
that a perturbation theory for 
the shifts in the Fermi occupation
function has a qualitative change 
to convergent behavior
when the hopping is still
relevant,
supports the notion
of incoherence directly.

\section{Conclusion}
We have calculated exactly the inter-Luttinger liquid
hopping rate to $O(t_{\perp}^2)$. Of great physical
relevance is the effective spectral function 
for interliquid hopping, $A_{12}(\omega)$. We have shown
that in a large region of Luttinger liquid parameter space
below $\alpha=1/2$ (the point where $t_{\perp}$ becomes a marginal
operator), $A_{12}(\omega)$ is too broad a function
to sustain coherent interliquid transport. 
Single particle
coherence is confined to the one-dimensional chains, in the 
sense that it is impossible to observe any interference effects
(beyond those observable for completely decoupled chains)
between histories which involve interliquid hopping.
Again, we emphasize that 
this is {\em not\/} the result of an irrelevant $t_{\perp}$:
the {\em coherence\/} is confined, but the {\em electrons\/} are not.

Our proposal is supported by a careful consideration of the
analytic properties of approximate single particle Green's functions.
Even though such approximations are uncontrolled, we find
no evidence in these calculations to suggest anything other than
that there can exist a phase of relevant, but incoherent, interliquid
transport. 
In fact, all of the results 
presented here indicate that 
motion of fermions transverse to the
chains
can be very different for different $\alpha$
(while still in the region of relevant $t_{\perp}$)
and support the idea that
the nature of the 
renormalization group instability
of the $t_{\perp}=0$ fixed point
can also change.
This gives further evidence 
for the existence of
a novel fixed point 
(one of ``confined coherence'') in which
transport in one or more (but not all) directions
is {\it intrinsically}
({\it i.e., in a pure system at zero temperature})
incoherent.

\samepage


%
%

\begin{figure}
\caption{The interliquid hopping spectral function for various values of 
$\alpha$. Here $\omega_l=(v_s-v)\Delta k$, $\omega_i=(v_c-v)\Delta k$
and $\omega_u$ is the ultraviolet cutoff of order $v/a$. The plots do not
include the weak power law cutoff dependent prefactors. The vertical
arrow is the $\alpha=0$ spectral function, 
$A_{12}(\omega)\propto \delta(\omega)$.}
\label{fig1}
\end{figure}

%
%


\begin{references}
\bibitem{footnote2} At energies of order the ultraviolet cutoff 
$\sim O(v/a)$ there are physically unimportant cutoff dependent corrections to 
$A_{12}^{\rm high}$.
\end{references}
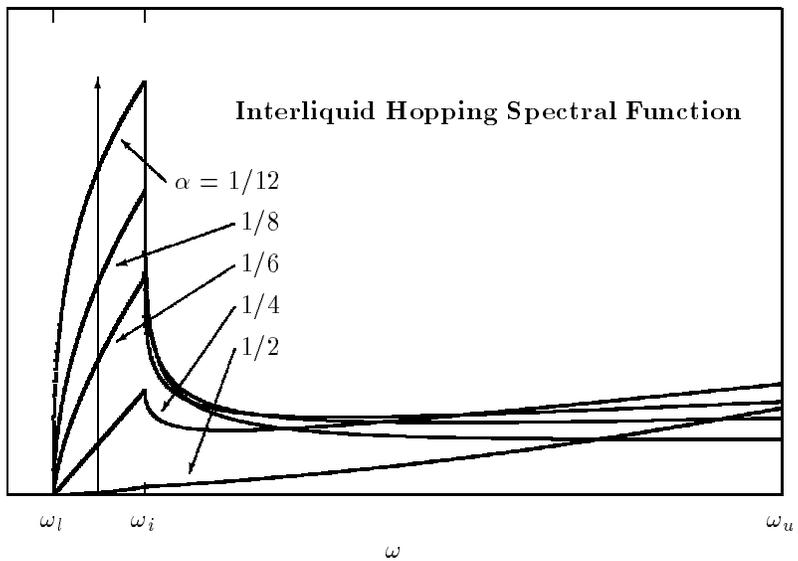
\end{document}